\documentclass[prd,preprintnumbers,amssymb,amsmath,12pt,nofootinbib,longbibliography]{revtex4}

\usepackage{graphicx}
\usepackage{subfigure}
\usepackage{amsmath}

\begin{document}

\title{$A_{FB}$ as a discovery tool for $Z^\prime$ bosons at the LHC}

\date{\today}

\author{Elena Accomando}%
 \email{E.Accomando@soton.ac.uk}
\affiliation{School of Physics and Astronomy, University of Southampton, Highfield, Southampton SO17 1BJ, UK}%
\author{Alexander Belyaev}%
 \email{A.Belyaev@soton.ac.uk}
\affiliation{School of Physics and Astronomy, University of Southampton, Highfield, Southampton SO17 1BJ, UK}%
\author{Juri Fiaschi}%
 \email{Juri.Fiaschi@soton.ac.uk}
 \affiliation{School of Physics and Astronomy, University of Southampton, Highfield, Southampton SO17 1BJ, UK}%
\author{Ken Mimasu}%
 \email{K.Mimasu@sussex.ac.uk}
\affiliation{School of Physics and Astronomy, University of Sussex, Falmer, Brighton, BN1 9RH, UK}%
\author{Stefano Moretti}%
 \email{S.Moretti@soton.ac.uk}
\affiliation{School of Physics and Astronomy, University of Southampton, Highfield, Southampton SO17 1BJ, UK}%
\author{Claire Shepherd-Themistocleous}%
 \email{claire.shepherd@stfc.ac.uk}
\affiliation{Particle Physics Department,
STFC, Rutherford Appleton Laboratory, Harwell Science and Innovation Campus,
Didcot, Oxfordshire, OX11 0QX, UK}%

\begin{abstract}
The Forward-Backward Asymmetry (AFB) in $Z^\prime$ physics is commonly only perceived as the observable which possibly allows one to interpret a $Z^\prime$ signal by distinguishing different models of such (heavy) spin-1 bosons.
In this article, we examine the potential of AFB in setting bounds on or even discovering a $Z^\prime$ at the Large Hadron Collider (LHC) and show that it might be a powerful tool for this purpose.
We analyze two different scenarios: $Z^\prime$s with a narrow and wide width, respectively. We find that in both cases AFB can complement the cross section in accessing $Z^\prime$ signals.
\end{abstract}

\maketitle

\section{Introduction}
Extra gauge bosons are present in many Beyond Standard Model (BSM) theories. From a phenomenological point of view, the simplest case is an extra $U(1)$ symmetry group in addition to the SM group. Using this approach we can study the three main classes of models that predict a $Z^\prime$: $E_6$, Generalized Left-Right (GLR) symmetric and Generalized Standard Model (GSM) \cite{Accomando:2010fz}.
All these scenarios predict rather narrow $Z^\prime$s ($\Gamma_{Z^\prime} / M_{Z^\prime} \sim 0.5 - 12 \%$).

Experimental searches optimized for such narrow resonances assume a very visible peak with a Breit-Wigner line-shape over the SM background, when looking at the invariant mass of the $Z^\prime$ decay products.
On the basis of this assumption, the 95\% Confidence Level (C.L.) upper bound on the cross section is derived and limits on the mass of the resonance are extracted within the above benchmark models.
Theoretical cross section predictions are usually calculated in Narrow Width Approximation (NWA), or possibly they might include Finite Width (FW)	 and interference effect. These can be taken into account in a model independent way, putting an appropriate cut on in the invariant mass spectrum \cite{Accomando:2013sfa}. 

However, there exist many scenarios where the NWA is not valid. Technicolor \cite{Belyaev:2008yj}, Composite Higgs Models \cite{Barducci:2012kk},
scenarios where the $Z^\prime$ couples differently to the first two fermion generations with respect to the third one \cite{Kim:2014afa, Malkawi:1999sa} or
where the new gauge sector mixes with the SM neutral one \cite{Altarelli:1989ff} are all frameworks where wide $Z^\prime$s are possible. Here,
the ratio $\Gamma_{Z^\prime}/M_{Z^\prime}$ can reach the 50\% value or more.

Experimental searches studying these 'effectively' non-resonant cases are essentially counting experiments: any integration over the overall invariant mass spectrum beyond  the control region seeks an excess of events spread over the SM background.
The analysis, even if improved using optimized kinematical cuts, still maintains some fragile aspects as it relies on the good understanding of the SM background. Indeed the BSM signal might not trivially interfere with the latter, affecting the $Z^\prime$ decay product invariant mass distribution also in the low mass region.
For this reason the detection of a wide resonance turns out to be quite problematic.

In this article we study the effects of the inclusion of another observable into the analysis of heavy neutral resonances: the Forward-Backward Asymmetry (AFB). We explore the complementary potential of AFB with respect to the 'bump' 
or `counting experiment' searches in both the narrow and broad $Z^\prime$ framework, respectively. Note that,
in current literature, this observable is usually adopted as a post-discovery tool to interpret the experimental evidence of a peaked signal and to possibly disentangle between different theoretical models that would predict it.
Our purpose is to show that AFB can be used not only for interpreting a possible discovery but also in the very same search process.
We focus on the $Z^\prime$ discovery golden channel search at the LHC, {\it i.e.}, the Drell-Yan (DY) process $pp\rightarrow l^+l^-$ with $l=e, \mu$.

The article is organized as follows. 
In sect. 2 we derive current and projected bounds for $Z^\prime$ model benchmarks for the LHC at 7, 8 and 13 TeV.
In sect. 3 we discuss the role of AFB  in the context of either narrow or wide resonance searches.
In sect. 4 we summarize and conclude.

\section{Bounds on the $Z^\prime$ mass}
In order to validate our analysis we reproduced current experimental limits obtained by, \textit{e.g.}, the CMS collaboration after the 7 and 8 TeV runs with about 20 $fb^{-1}$ of luminosity, assuming the NWA \cite{Khachatryan:2014fba}.
These limits are computed through the ratio $R_\sigma = \sigma (pp\rightarrow Z^\prime\rightarrow l^+l^-)/\sigma (pp\rightarrow Z, \gamma \rightarrow l^+l^-)$.% with $l = e, \mu$. The use of this ratio $R_\sigma$ in fact cancels the uncertainty in the integrated luminosity and reduces the dependence on the experimental acceptance and  trigger efficiency. Here,
$R_\sigma$ has been calculated at the Next-to-Next-to-Leading Order (NNLO) in QCD using the WZPROD program \cite{Hamberg:1990np,vanNeerven:1991gh,ZWPROD} (which we have adapted for $Z^\prime$ models \cite{Accomando:2010fz} and the CTEQ6.6 package \cite{Kretzer:2003it}). 

The resulting exclusion limits we compute include FW and interference effects. The values we obtain are summarized in Tab. \ref{tab:events_8tev}: they match the reported limits by CMS for the benchmark models GSM-SSM and $E_6-\chi$ within the accuracy of 1-2 \% (except for the $Q$ model which predicts a slightly wider resonance and thus the discrepancy with the CMS results is around 5\%).

\begin{table}
  \caption{Bounds on the $Z^\prime$ mass derived from the latest direct searches performed by CMS at the 7 and 8 TeV LHC with integrated luminosity $L=20 fb^{-1}$.}
  \label{tab:events_8tev}
  \begin{tabular}{|c||c|c|c|c|c|c|c|c|c|c|c|c|c|}
    \hline
      Class & \multicolumn{6}{c|}{$E_6$} & \multicolumn{4}{c|}{GLR} & \multicolumn{3}{c|}{GSM} \\
    \hline
      $U^\prime (1)$ Models  & $\chi$ & $\phi$ & $\eta$ & $S$ & $I$ & $N$ & $R$ & $B-L$ & $LR$ & $Y$ & $SSM$ & $T_{3L}$ & $Q$ \\
    \hline
      $M_{Z^\prime}$ [GeV] & 2700 & 2560 & 2620 & 2640 & 2600 & 2570 & 3040 & 2950 & 2765 & 3260 & 2900 & 3135 & 3720\\
    \hline
  \end{tabular}
\end{table}

It is worth stressing that, in the context of narrow resonance searches, CMS adopted a dedicated cut on the invariant mass of the di-lepton pairs: $|M_{l\bar l}-M_{Z^\prime}|\le 0.05\times E_{\rm LHC}$ where $E_{\rm LHC}$ is the collider energy. 
This cut was designed so that the error in neglecting the (model-dependent) FW and interference effects (between $\gamma , Z, Z^\prime$) are kept below $O(10\% )$ for all models and  the full range of allowed $Z^\prime$ masses under study, thus following the recommendations of \cite{Accomando:2013sfa}.

After having verified the reliability of our code, we have been able to project future discovery and exclusion limits for the next run of the LHC at 13 TeV and with a luminosity of 300 $fb^{-1}$. In both cases we have taken into account the published acceptance $\times$ efficiency corrections and a Poisson statistic approach has been used for computing the significance of the signal.
Requiring for the latter a significance of 2 for exclusion and 5 for discovery, we obtain the results summarized in Tab. \ref{tab:events_13tev}.

\begin{table}
  \caption{Projection of discovery limits (first row) and exclusion limits (second row) on the $Z^\prime$ mass from direct searches at the forthcoming Run II of the LHC at 13 TeV. We assume $L=300 ~fb^{-1}$.}
  \label{tab:events_13tev}
  \begin{tabular}{|c||c|c|c|c|c|c|c|c|c|c|c|c|c|}
    \hline
      Class & \multicolumn{6}{c|}{$E_6$} & \multicolumn{4}{c|}{GLR} & \multicolumn{3}{c|}{GSM} \\
    \hline
      $U^\prime (1)$ Models  & $\chi$ & $\phi$ & $\eta$ & $S$ & $I$ & $N$ & $R$ & $B-L$ & $LR$ & $Y$ & $SSM$ & $T_{3L}$ & $Q$ \\
    \hline
      $M_{Z^\prime}$ [GeV] & 4535 & 4270 & 4385 & 4405 & 4325 & 4290 & 5175 & 5005 & 4655 & 5585 & 4950 & 5340 & 6360 \\
    \hline
      $M_{Z^\prime}$ [GeV] & 5330 & 5150 & 5275 & 5150 & 5055 & 5125 & 6020 & 5855 & 5495 & 6435 & 5750 & 6180 & 8835 \\
    \hline
  \end{tabular}
\end{table}

\section{The role of AFB in $Z^\prime$ searches: narrow and wide heavy resonances}

We define AFB as follows:
\begin{equation}
\label{eq:AFB}
\frac{d\sigma}{d\cos\theta_l^*}\propto \sum_{spin,col}\left|\sum_i\mathcal{M}_i\right|^2=\frac{\hat{s}^2}{3}\sum_{i,j}|P^*_iP_j|[(1+\cos^2\theta_l^*)C^{ij}_S+2\cos\theta_l^* C^{ij}_A] 
\end{equation}
where $\theta_l^*$ is the lepton angle with respect to the quark direction in the di-lepton Centre-of-Mass  (CM) frame, which can be derived from the measured four-momenta of the di-lepton system in the laboratory frame. 
The AFB is indeed given by the coefficient of the contribution to the angular distribution linear in $\cos\theta_l^*$. In Eq. (\ref{eq:AFB}), $\sqrt{\hat{s}}$ is the invariant mass of the di-lepton system and $P_i$ and $P_j$ are the propagators of the gauge bosons involved in the process. 
At the tree-level, DY production of charged lepton pairs is mediated by three gauge bosons: the SM photon and $Z$-boson and the hypothetical $Z^\prime$. These three vector boson exchanges all participate in the matrix element squared. The interferences amongst these three particles  have to be take into account properly.
Finally, the factors $C_{S}^{ij}$ and $C_{A}^{ij}$ in the angular distribution given in Eq. (\ref{eq:AFB}) are the parity symmetric and anti-symmetric coefficients which are functions of the chiral quark and lepton couplings, $q_{L/R}^i$ and $e_{L/R}^i$, to the $i$-boson with $i=\{\gamma, Z, Z^\prime\}$:

\begin{align}
\label{eqn:def_C}
C_{S}^{ij}&=(q_L^i q_L^j+ q_R^i q_R^j)(e_L^i e_L^j+ e_R^i e_R^j), \\           
C_{A}^{ij}&=(q_L^i q_L^j- q_R^i q_R^j)(e_L^i e_L^j- e_R^i e_R^j).   
\end{align}
\noindent

Looking at these expressions it is clear that the analysis of AFB can give us complementary informations with respect to the cross section distribution (which is proportional to the sum of the squared chilar couplings)
about  the couplings between the $Z^\prime$ and the fermions.
This feature has motivated several authors to study the potential of AFB  in interpreting a possible $Z^\prime$ discovery obtained in the usual cross section hunt as, {\it e.g.},  in Refs. \cite{Langacker:1984dc, Carena:2004xs, Petriello:2008zr, Rizzo:2009pu}. 

The AFB is obtained by integrating the lepton angular distribution forward and backward with respect to the quark direction.
As in $pp$ collisions the original quark direction is not known, one has to extract it from the kinematics of the di-lepton system. In this analysis, we follow the criteria of Ref. \cite{Dittmar:1996my} and simulate the quark direction from the boost of the di-lepton system with respect to the beam axis ($z$-axis). 
This strategy is motivated by the fact that at the LHC the di-lepton events at high invariant mass come from the annihilation of either valence quarks with sea antiquarks or sea quarks with sea antiquarks. As the valence quarks carry away, on average, a much larger fraction of the proton momentum than the sea antiquarks, the boost direction of the di-lepton system should give a good approximation of the quark direction. 
A leptonic forward-backward asymmetry can thus be expected with respect to the boost direction. In contrast, the subleading number of di-lepton events which originate from the annihilation of quark-antiquark pairs from the sea must be symmetric.

As a measure of the boost, we define the di-lepton rapidity $y_{l\bar{l}}=\frac{1}{2} {\rm{ln}}\left [\frac{E+P_z}{E-P_z}\right ]$,
where $E$ and $P_z$ are the energy and the longitudinal momentum of the di-lepton system. We identify the quark direction through the sign of $y_{l\bar{l}}$. In this way, one can define the `reconstructed'  AFB, from now on called $A_{FB}^\ast$. 
Namely, we have defined $A_{FB}^\ast$ using the $\theta_l^\ast$ reconstructed angle, which is the angle between the final state lepton and the incoming quark direction in the CM of the di-lepton system.

In the following we are going to show the impact of AFB on the significance of the signal. For this purpose we give the general definition of significance $\alpha$ for a generic observable:
\vspace*{-0.1truecm}
\begin{equation}
\alpha=\frac{|O_1-O_2|}{\sqrt{\delta O^2_1+\delta O^2_2}}.
\end{equation}

where the $O_i$s ($i=1,2$) are the value of the observable in two hypothesis scenarios with uncertainty $\delta O_i$. In the case of AFB we will use the statistical uncertainty given by:
\begin{equation}
\delta A_{FB} =\sqrt{\frac{4}{\mathcal{L}}\frac{\sigma_{F}\sigma_{B}}{(\sigma_{F}+\sigma_{B})^3}} = 
\sqrt{\frac{(1-A_{FB}^2)}{\sigma\mathcal{L}}} = \sqrt{\frac{(1-A_{FB}^2)}{N}},
\end{equation}

where $\mathcal{L}$ is the integrated luminosity and $N$ the total number of events. One can thus see that the significance is proportional to the root of the total number of events. 
This mean that the imposition of a stringent cut on the boost variable, $y_{l\bar{l}}$, in spite of guiding the AFB spectrum towards its true line shape, will decrease the statistics and, by consequence, the resulting significance of the signal.

With this background in mind we are going to show how AFB can be used also as a powerful tool to search for new physics.

\subsection{Narrow heavy resonances}

\begin{figure}[t]
\centering
\subfigure[]{
\includegraphics[width=0.47\textwidth]{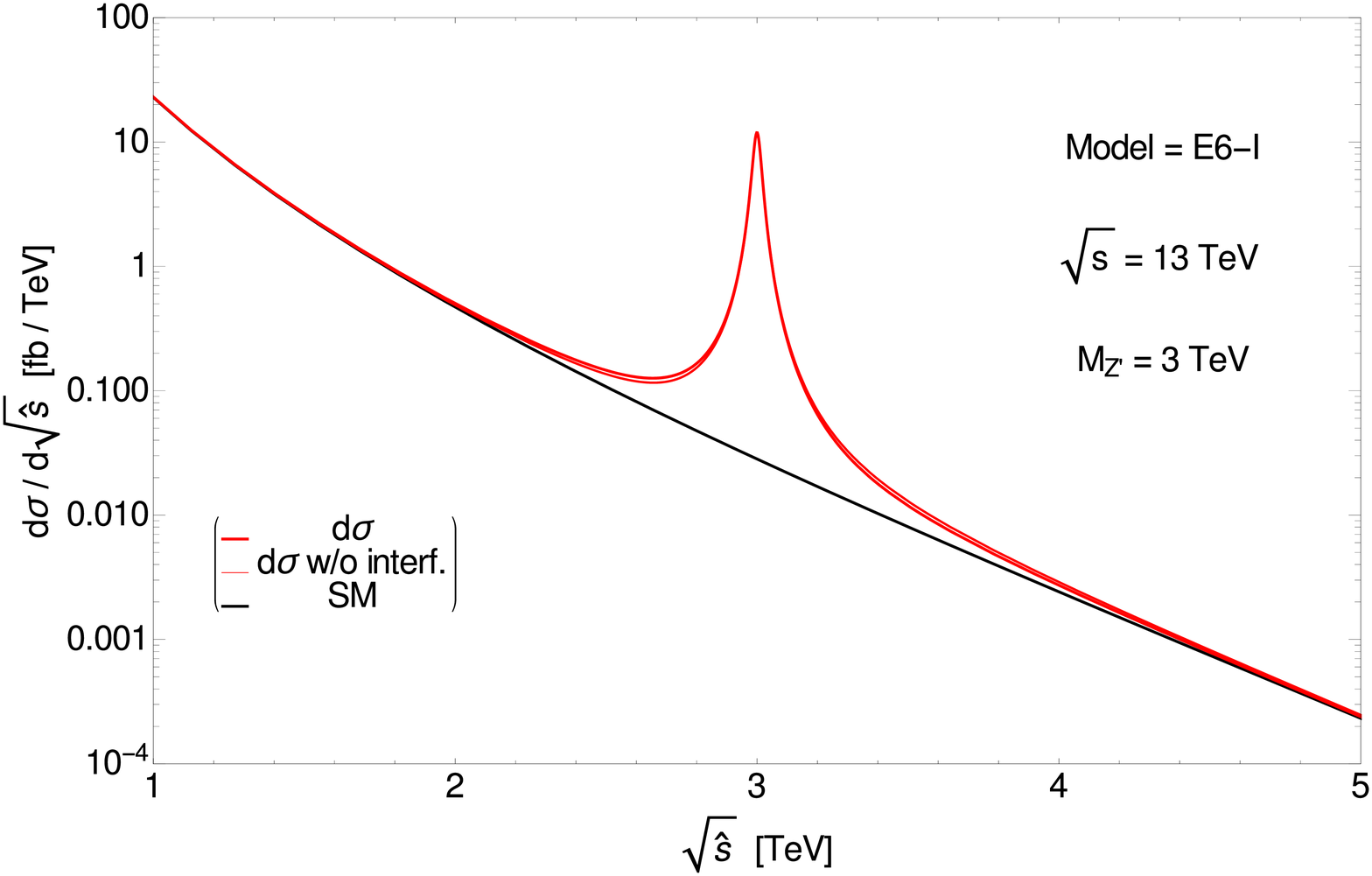}
\label{fig:sigma_I_analytic}
}
\subfigure[]{
\includegraphics[width=0.47\textwidth]{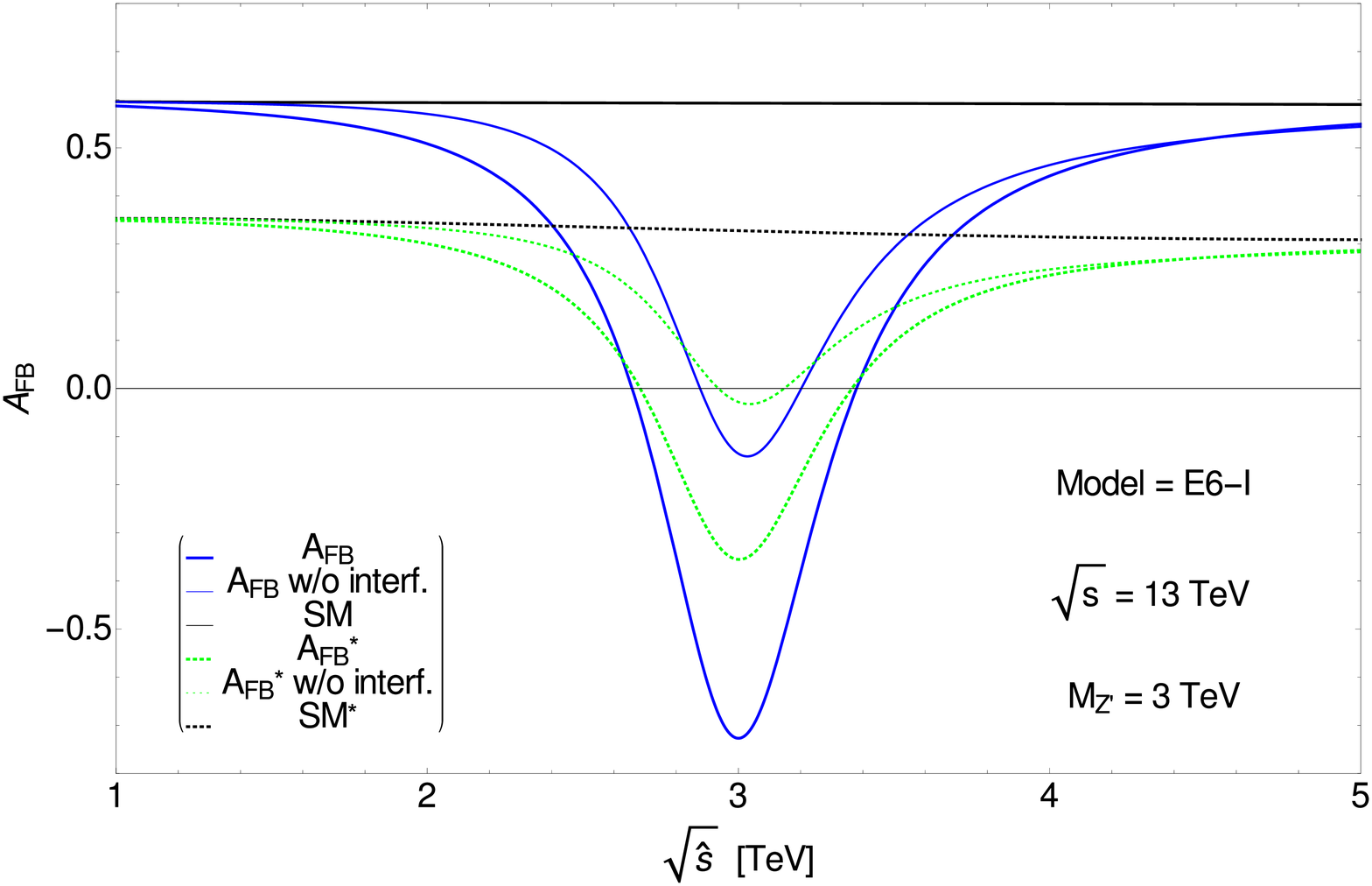}
\label{fig:afb_I_analytic}
}
\subfigure[]{
\includegraphics[width=0.47\textwidth]{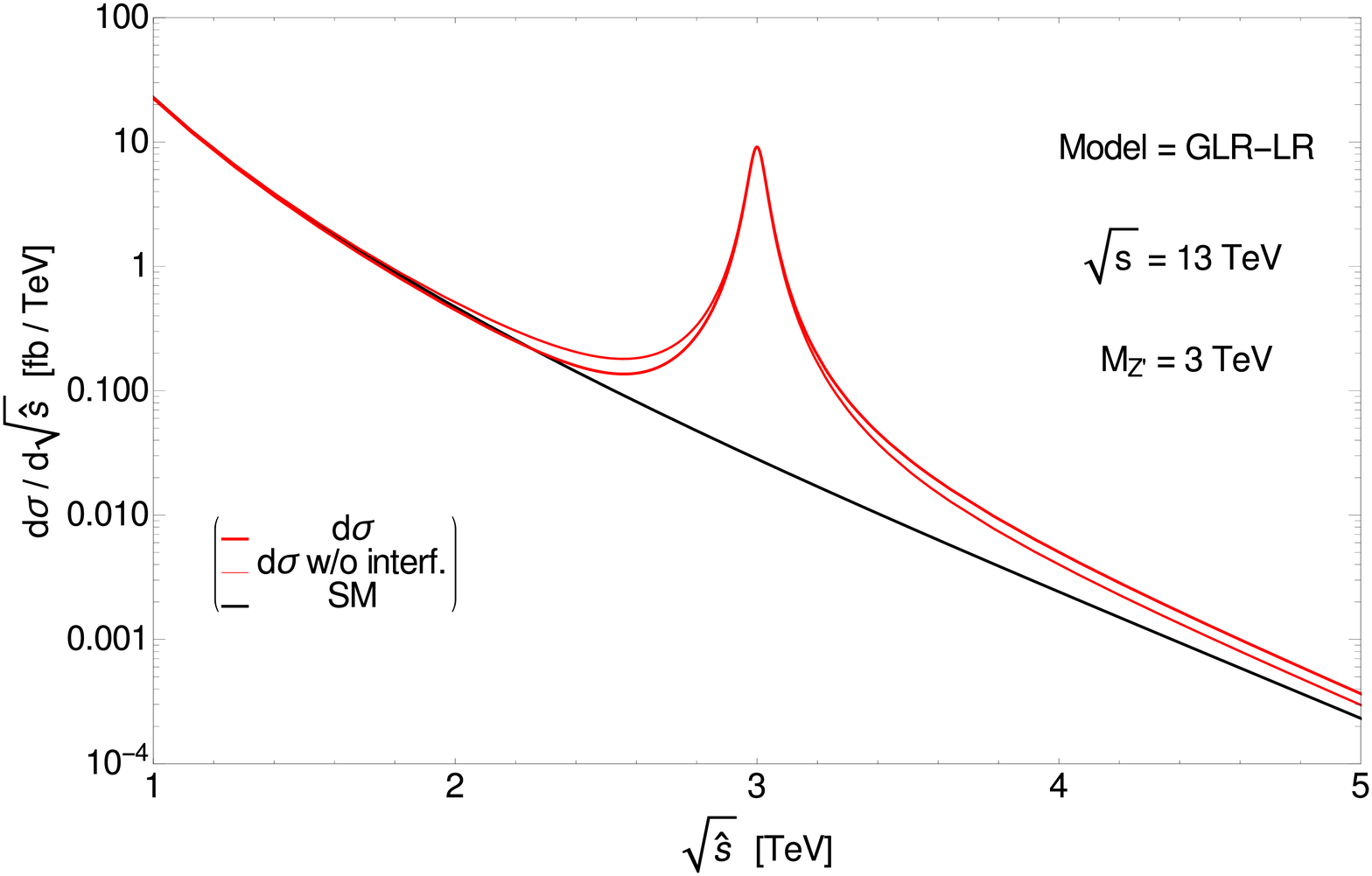}
\label{fig:sigma_LR_analytic}
}
\subfigure[]{
\includegraphics[width=0.47\textwidth]{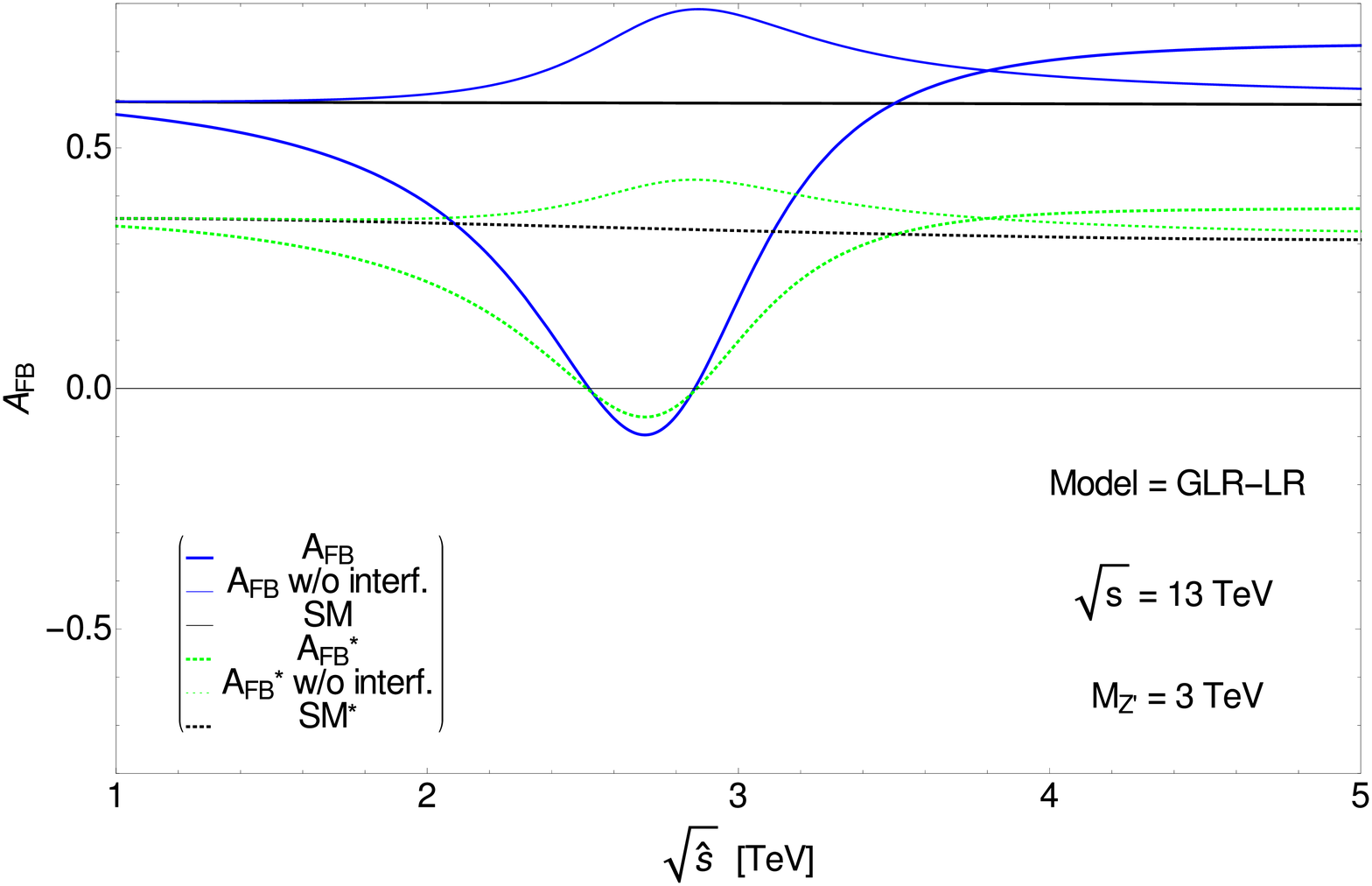}
\label{fig:afb_LR_analytic}
}
\caption{
\subref{fig:sigma_I_analytic} Hypothetic signal in the cross section distribution produced by a $Z^\prime$ with mass $M_ {Z^\prime}$ = 3 TeV, as predicted by the $E_6$-I model, at the LHC at $\sqrt{s}$=13 TeV. 
\subref{fig:afb_I_analytic} Hypothetic signal in the $A_{FB}^*$ distribution produced by a $Z^\prime$ with mass $M_ {Z^\prime}$ = 3 TeV, as predicted by the $E_6$-I model, at the LHC at $\sqrt{s}$=13 TeV. No cut on the di-lepton rapidity is imposed: $|y_{l\bar{l}}|\ge 0$. 
\subref{fig:sigma_LR_analytic} Same as plot (a) within the GLR-LR model.
\subref{fig:afb_LR_analytic} Same as plot (b) within the GLR-LR model.
}
\label{fig:Narrow_models}
\end{figure}

We start comparing the shape of the AFB distribution as a function of the di-lepton invariant mass $\sqrt{\hat{s}}$ with the differential cross section distribution.
We are showing here two interesting cases: the $E_6$-I (Figs. \ref{fig:sigma_I_analytic} and \ref{fig:afb_I_analytic}) and GLR-LR (Figs. \ref{fig:sigma_LR_analytic} and \ref{fig:afb_LR_analytic}) models.

As one can see, the role played by the interference is extremely important (also when reconstructed). In the $E_6$-I case the AFB peak is heavily accentuated, while in the GLR-LR case the peak is shifted to a lower value in the invariant mass distribution.
In contrast, the cross section distribution is almost interference free if the $|M_{l\bar l}-M_{Z^\prime}|\le 0.05\times E_{\rm{LHC}}$ cut is imposed. In interpreting the experimental data coming from AFB measurements instead it is mandatory to include the interference independently on any kinematical cut.

\begin{figure}[t]
\centering
\subfigure[]{
\includegraphics[width=0.47\textwidth]{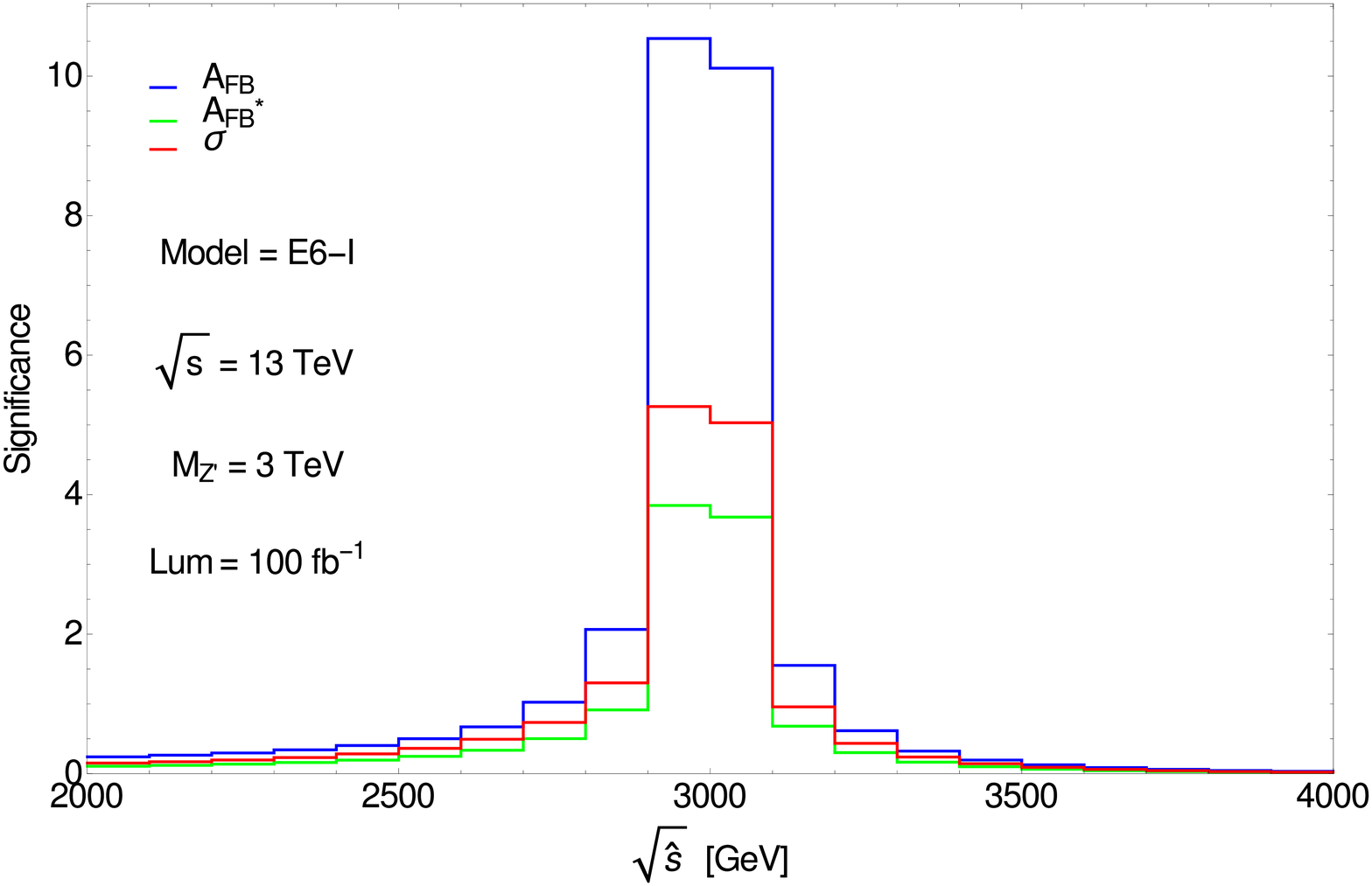}
\label{fig:significance_I}
}
\subfigure[]{
\includegraphics[width=0.47\textwidth]{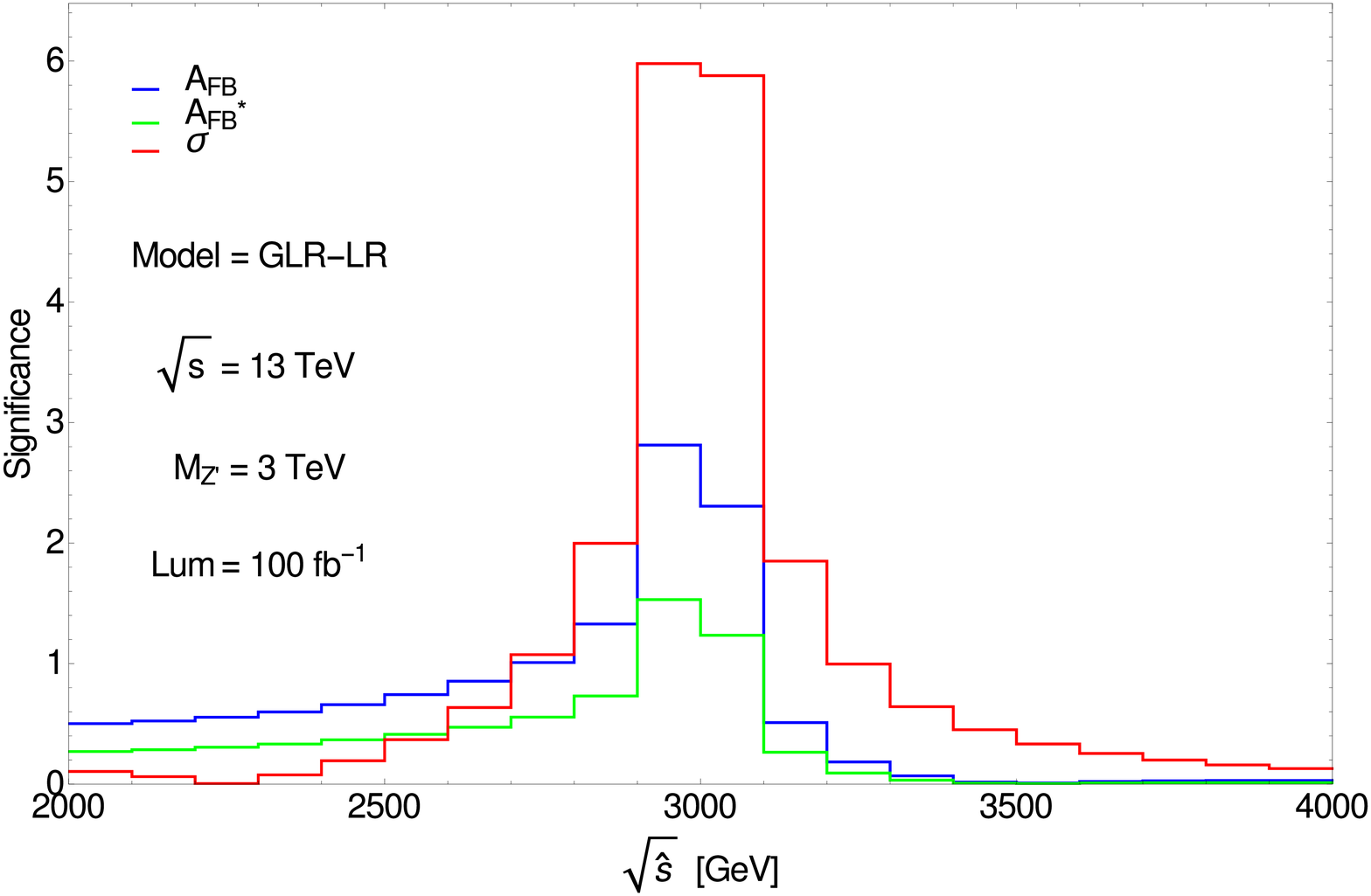}
\label{fig:significance_LR}
}
\caption{
\subref{fig:significance_I} Binned significance of an hypothetic signal produced by a $Z^\prime$ with mass $M_ {Z^\prime}$ = 3 TeV, as predicted by the $E_6$-I model, at the LHC at $\sqrt{s}=13$ TeV and $\mathcal{L}=100 fb^{-1}$, in the three observables: cross section, $A_{FB}$ and $A_{FB}^*$.
\subref{fig:significance_LR} Same as (a) for the GLR-LR model.
}
\label{fig:Narrow_models_significance}
\end{figure}

In terms of significance of the AFB signal (Fig. \ref{fig:Narrow_models_significance}), we obtain that for the $E_6$-I case the  peak leads to a significance which is comparable with what we get from the bump in the cross section, even after
reconstruction.
Thus, it can be used as a very valid alternative as the AFB observable is very reliable in terms of systematic uncertainties: since it comes from the ratio of scross sections, strong cancellations happen between the uncertainties on the forward and backward cross sections, upon taking into account their mutual correlations.
In the GLR-LR case instead the interference effects shift the AFB peak to a lower invariant mass region, as mentioned above, which might  lead to an early hint of the presence of new physics, {\it i.e.}, even before the $Z^\prime$ pole is
reached.

\subsection{Wide heavy resonances}
Here, we discuss the role of  $A_{FB}^*$ in searches for a new $Z^\prime$ characterized by a large width. 
Such a heavy and wide particle is predicted by various models.
A benchmark scenario for experimental analyses is the wide version of the SSM described in Ref. \cite{Altarelli:1989ff}.
The proposal is to have a heavy copy of the SM neutral gauge boson $Z$, with same couplings to ordinary matter and SM gauge bosons. 
Owing to the $Z^\prime$ decay into SM charged gauge bosons, whose rate grows with the third power of the $Z^\prime$ mass, the total width of the new heavy particle can be quite large: $\Gamma_{Z^\prime}/M_{Z^\prime}\simeq 50\%$ and above. 

\begin{figure}[t]
\centering
\subfigure[]{
\includegraphics[width=0.47\textwidth]{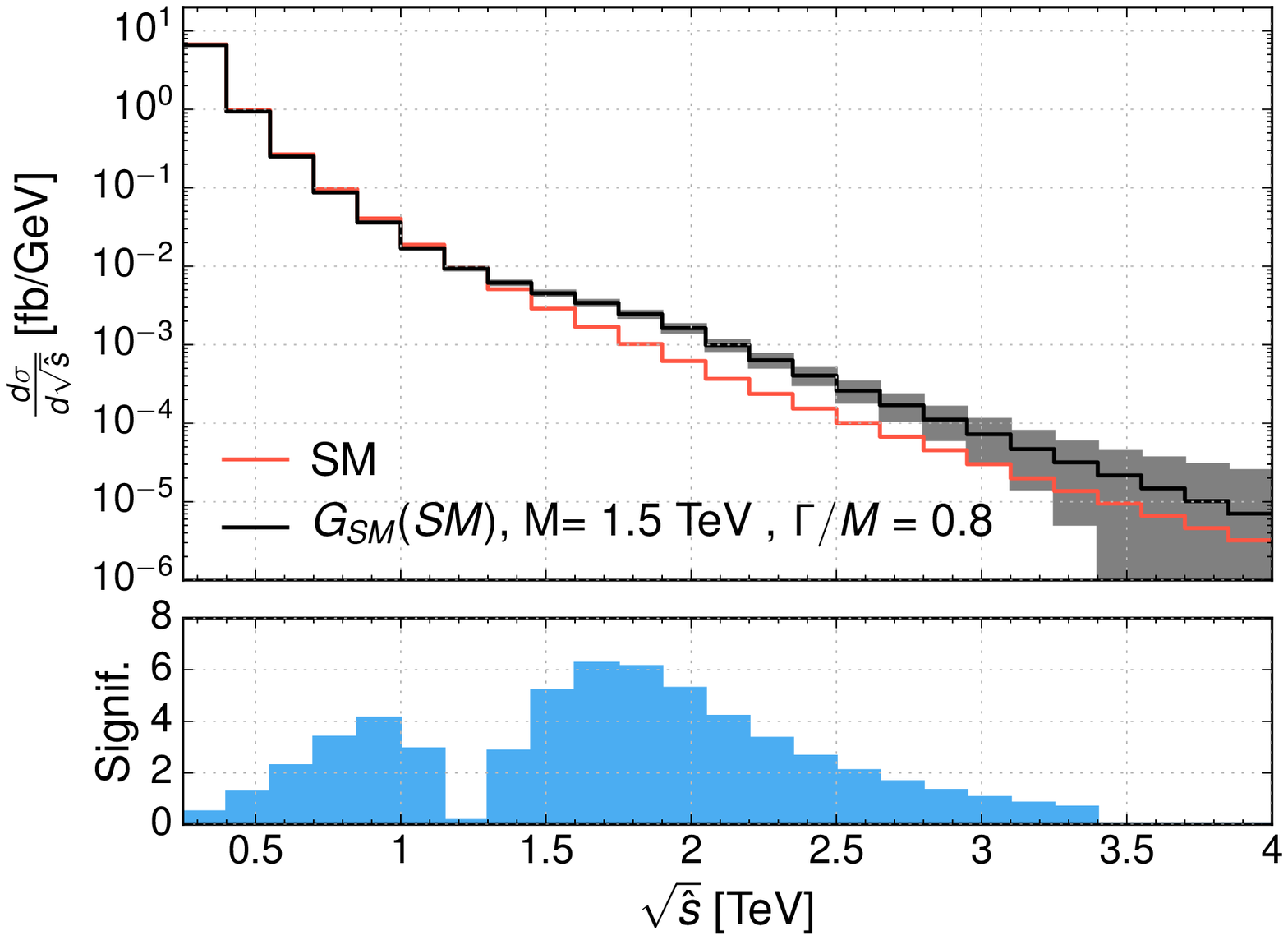}
\label{fig:sigma_gsmwide_realistic}
}
\subfigure[]{
\includegraphics[width=0.47\textwidth]{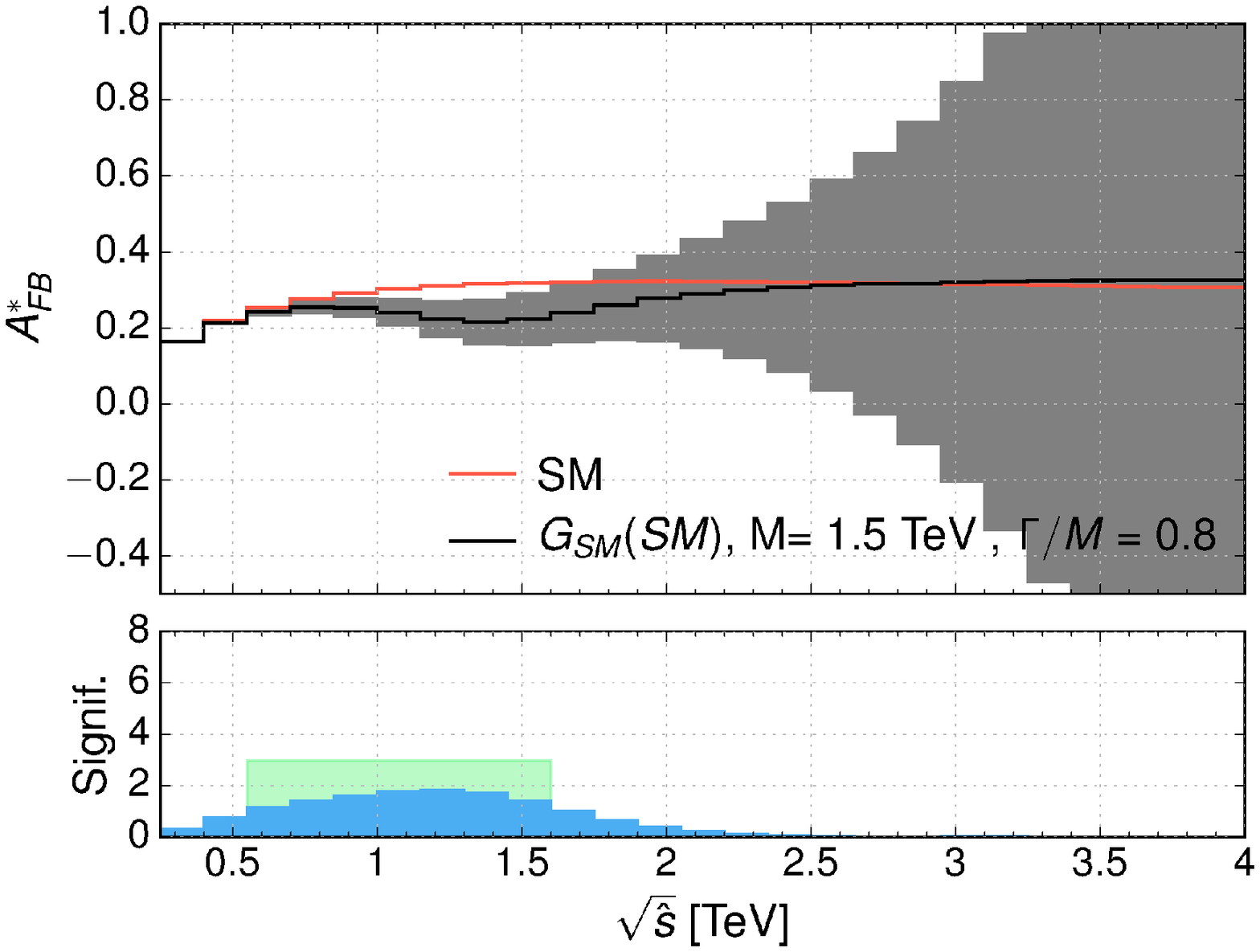}
\label{fig:AFB_gsmwide_realistic}
}
\caption{
\subref{fig:sigma_gsmwide_realistic} Binned differential cross section as a function of the di-lepton invariant mass as predicted by the GSM-SM model for a $Z^\prime$ with mass $M_ {Z^\prime}$ = 1.5 TeV and $\Gamma_{Z^\prime}/M_{Z^\prime}= 80 \%$. 
Error bars are included. The results are for the LHC at $\sqrt{s}$=13 TeV and $\mathcal{L}=300 fb^{-1}$. Acceptance cuts are included ($p_T > 25$ GeV and $y_l < 2.5$).
\subref{fig:AFB_gsmwide_realistic} Same as plot (a) for the $A_{FB}^*$ distribution.}
\label{fig:gsmmodels_realistic}
\end{figure}

In this case, the invariant mass distribution of the two final state leptons does not show in the cross section a resonant (or peaking) structure around the physical mass of the $Z^\prime$ standing sharply over a smooth background, but just a broad shoulder spread over the SM background. 
This result is plotted in Fig. \ref{fig:gsmmodels_realistic}a, where we consider a $Z^\prime$ with mass $M_ {Z^\prime}$ = 1.5 TeV and width $\Gamma_{Z^\prime}/M_{Z^\prime}= 80 \%$. 
The line shape of the resonance is not well defined but the $A_{FB}^*$ observable could help to interpret a possible excess of events and it is shown in Fig. \ref{fig:gsmmodels_realistic}b. 
From the significance plots below one can see that the $A_{FB}^*$ shape could be visible at the 2$\sigma$ level in a region
where the significance from the cross section is decreasing.

The experimental method based on the counting experiment is based on the assumption that the control region is new physics free. But, this is not the case for wide $Z^\prime$s. 
In these scenarios, the interference between the extra $Z^\prime$ and the SM $\gamma , Z$ is so sizable that it can invade the control region. 
If not correctly interpreted, these interference effects could induce one to underestimate the SM background with the consequence of overestimating the extracted mass bounds.  
Having all these uncertainties to deal with, the support of a second observable like AFB is thus crucial for wide $Z^\prime$ searches.

\section{Conclusions}
In this paper we have considered the scope of using AFB in 
$Z^\prime$ searches at the LHC in the neutral DY channel.
Such a variable has traditionally been used for diagnostic purposes in presence of a potential signal previously established
through a standard resonance search via the cross section. However, based on the observation that it is affected 
by systematics less than cross sections (being a ratio of the latter), we
have studied the possibility of using AFB as a search tool for a variety of $Z^\prime$ models, $E_6$, GLR, GSM,
embedding either a narrow or wide resonance. The focus was on determining whether such a resonance could be sufficiently wide and/or weakly coupled such that a normal resonance search may not fully identify it and, further, whether the 
AFB could then provide a signal of comparable or higher significance to complement or even
surpass the scope of more traditional analyses. 

We have found promising results. In the case of narrow width $Z^\prime$s, we have proven that 
the significance of AFB based searches can be comparable with the usual bump search.  Further, we have
emphasized the fact that the AFB distribution mapped in di-lepton invariant mass can present features amenable
to experimental investigation not only in the peak region but also significantly away from the latter. 
In the case of
wide $Z^\prime$, the AFB search could have a better sensitivity than the cross section studies 
thanks to a more peculiar line-shape.
In essence, here, AFB in specific regions of the invariant mass of the reconstructed $Z^\prime$ could be sensitive to broad resonances much more than the cross section, wherein the broad distribution of the signal seemingly merges with the background.\\

\noindent
{\bf Acknowledgements}~
We are grateful to Patrik Svantesson for stimulating discussions.
This work is supported  by the Science and Technology Facilities Council, grant number  ST/L000296/1.
All authors acknowledge partial financial support through the NExT Institute.
\vspace*{-0.25truecm}

\bibliography{references}

\end{document}